\documentclass[twocolumn,showpacs,preprintnumbers,amsmath,amssymb]{revtex4}  
  
\usepackage{graphicx}
\usepackage{dcolumn}
\usepackage{bm}
\def\gap{\;\rlap{\lower 2.5pt  
 \hbox{$\sim$}}\raise 1.5pt\hbox{$>$}\;}  
\def\lap{\;\rlap{\lower 2.5pt  
   \hbox{$\sim$}}\raise 1.5pt\hbox{$<$}\;}  
\def\gsim{\;\rlap{\lower 2.5pt  
 \hbox{$\sim$}}\raise 1.5pt\hbox{$>$}\;}  
\def\lsim{\;\rlap{\lower 2.5pt  
   \hbox{$\sim$}}\raise 1.5pt\hbox{$<$}\;}  
\def\msun{{M_\odot}}  
\def\rsun{{R_\odot}}

\def\cm{{\rm\,cm}}  
  
\def\kpc{{\rm\,kpc}}  
\def\pc{{\rm\,pc}}

\def\GeV{{\rm\,GeV}}  
\def\TeV{{\rm\,TeV}}  
\def\sec{{\rm\,s}}  
\def\sr{{\rm\,sr}}

\def\MeV{{\rm\,MeV}}

\begin{document}  
  
\preprint{PRL/V0}  
  
\title{Difficulty of detecting minihalos via $\gamma$-rays from dark matter 
annihilation}
  
\author{Lidia Pieri$^1$, Enzo Branchini$^2$ and Stefan Hofmann$^3$}  
\affiliation{  
$^1$ {Department of Physics, Stockholm University, AlbaNova University Center, 
SE-10691 Stockholm, Sweden}\\  
$^2$ Department of Physics, Universit\`a di Roma Tre, Via della Vasca Navale 84, I-00146 Rome, Italy\\  
$^3$ Perimeter Institute for Theoretical Physics, Waterloo, Ontario, 
N2L 2Y5, Canada\\  
}  
\date{\today}
  
\begin{abstract}  
Analytical calculations and recent numerical experiments have shown   
that a sizable of the mass in our Galaxy is in a form of clumpy, 
virialized substructures that, according to \cite{dms:05}, 
can be as light as $10^{-6} \msun$.  
In this work we estimate the gamma-rays flux expected from dark matter  
annihilation occurring within these minihalos, under the hypothesis  
that the bulk of dark matter is composed by neutralinos.  
We generate mock sky maps showing the angular distribution of  
the expected gamma-ray signal. We compare them with  
the sensitivities of satellite-borne experiments such as GLAST and   
find that a possible detection of minihalos is indeed very  
challenging. 
\end{abstract}  
  
  
  
\maketitle   
  
Analyses of the anisotropies in the cosmic microwave background 
radiation \cite{spe:03}  
find that the matter density content of the universe 
is approximately six times larger  
than the baryonic one, in agreement with
the observed abundance of light 
elements \cite{tyt:00} and the matter power spectrum inferred from galaxy 
redshift surveys \cite{teg:03}.
These observations involve very different physics, but unambiguously indicate  
that the Universe contains a significant amount of non-baryonic 
cold dark matter (CDM).  
The nature of CDM is presently unknown. However,   
weakly interacting massive particles (WIMPs) are regarded as generic 
particle candidates for CDM, as they naturally arise in extensions 
of the standard model of particle physics.
WIMPs are a particularly attractive CDM candidate, since the average 
mass density  of a stable relic from the electroweak scale is 
expected to be close to critical \cite{dim:90}.  
A further exciting feature of WIMPs is that they can be investigated 
in ongoing and future astrophysical direct \cite{dru:86} and indirect 
searches \cite{lbe:00} and in upcoming laboratory experiments.   
  
Indirect CDM searches focus on measuring the diffuse flux of CDM annihilation  
products $\Phi_\gamma \equiv \Phi^{\rm SUSY} \times \Phi^{\rm cosmo}$ 
\cite{fps:04}.  
The particle physics dependence embedded in $\Phi^{\rm SUSY}$ involves 
the WIMP's annihilation cross-section and branching ratios as well as 
the final photon energy spectrum and 
can be calculated from the underlying WIMP field theory.  
In this work we assume the nearly mass independent best value for the 
annihilation cross section of 
$2 \times 10^{-26} \cm^3 \sec^{-1}$, which represents an upper bound 
still allowing the neutralinos to be the dominant CDM component. 
The geometry dependence of the flux is contained in  $\Phi^{\rm cosmo}$ that  
is found by integrating the square of the neutralino mass density along the   
line--of--sight and therefore is very sensitive to the presence of virialized  
structures representing maxima of the DM density field.  
In this work we focus on the annihilation signal produced within our 
Galaxy under the hypothesis, supported by several high resolution 
numerical experiments, that part of its mass is in the form 
of subhalos.  
  
CDM models are characterized by their excess power on small scales   
that leads to a logarithmic divergence of the linear density contrast 
at large wavenumbers $\Delta \equiv \delta\rho/\rho \propto {\rm ln}(k)$.    
Recent analytical calculations have proved that $\Delta$ shows exponential 
damping for $k > k_{\rm cut} = {\cal O}(1)/{\rm pc}$.   
The cut-off scale $k_{\rm cut}$ is given by viscous (collisional) processes   
before and free (collisionless) streaming after kinetic decoupling at   
$T_{\rm kd}={\cal O}(10) \MeV$, both leading to exponential damping of 
the linear CDM density contrast \cite{hss:01,ghs:04,ghs:05}.  During 
the linear regime most of the statistical power   
typically goes in density contrasts $\Delta$ with $k\sim k_{\rm cut}$.  
So $k_{\rm cut}$ sets also the typical scale for the first halos.
that typically form at a redshift $z_{\rm n} = 60 \pm 10$ 
(for the best fit WMAP matter density) 
\cite{ghs:04,ghs:05}, when the 
mass variance $\sigma = 1$ on a comoving length scale $R={\cal O}(1) {\rm pc}$.    
  
The fate of these early minihalos in the non linear regime when they 
typically merge into larger halos, representing higher levels in the 
hierarchy, can only be followed through numerical experiments. 
The limited dynamical range 
explored with currently available  
N-body codes makes it very challenging to perform numerical experiments 
with a resolution as high as $ k_{\rm cut}^{-1}$ on a computational 
box large enough to guarantee statistical significance.  
Yet, \cite{dms:05} have recently simulated hierarchical 
clustering in CDM cosmologies up to a dynamical resolution 
$k_{\rm res} \sim k_{\rm cut}$ in a small high resolution patch  
which is nested within a hierarchy of larger low resolution regions.   
They find a steep halo mass function ${\rm d}n(M) /{\rm d ln}(M) 
\propto M^{-1} {\rm exp}[-(M/M_{\rm cut})^{-\frac{2}{3}}]$,   
with $M_{\rm cut} = 5.7 \times 10^{-6} h^{-1} \msun \propto k_{\rm cut}^{\; -3}$ 
in the mass range  $[10^{-6}, 10^{-4}]\msun$ whose extrapolation fits well
that of Galactic halos \cite{reed:03}.

Since \cite{dms:05} stopped their experiment at $z=26$
and probed a small volume with a limited dynamical range,
no general consensus exists on whether these early structures 
survived within massive galactic halos until today and what their mass 
function and distribution is.
Indeed, although their mass concentration is probably large enough for them  
to survive the gravitational tides experienced in the external region
of the massive host halo  (see, however \cite{Berezinsky:03}), they could be 
torn into tidal stream by close encounters with individual stars in the galaxy  
\cite{zh:05}. These issues could only be settled by simulating the
gravitational clustering out to $z=0$ on a volume large enough to overlap
the results with those of other, high resolution numerical experiments 
of Galaxy size halos (\cite{sto:03,dms:04}), which is beyond current 
numerical limitations. Yet, \cite{dms:05,otn:05} have suggested that 
these minihalos might significantly contribute to the total annihilation
flux in our Galaxy.

In this work we compute the expected $\gamma$-ray flux produced by a 
population of sub-Galactic halos under the optimistic hypotheses
of \cite{dms:05} that all minihalos of masses $\ge 10^{-6} \msun$
that 
survived gravitational tides populate 
the Galactic halo at $z=0$, trace its mass, share the same
self-similar cuspy density profile and have a steep mass function $\propto M^{-1}$
up to $\ge 10^{10} \msun$. 
It is worth stressing that our predictions represent an upper 
limit to the expected photon 
flux and to its experimental detectability that we will investigate   
for the case of a satellite-borne experiment like GLAST \cite{GLAST}.   

Our analysis extends those of \cite{cm:00,pb:04,pmp:04,sto:03,savvas:04} since
we consider subhalos much lighter than $10^6  \msun$ and since we use
Monte Carlo techniques to explicitly account for the contribution 
of very nearby minihalos to the total annihilation flux.  
The relevant properties of our minihalo population are:
a steep mass function ${\rm d}n(M) /{\rm d ln}(M) \propto M^{-1}$
in the range $[10^{-6}, 10^{10}]\msun$,
a spatial distribution that trace the smooth mass component 
in our Galaxy and a NFW mass density profile \cite{navarro:97} 
for both the MW and for all the subhalos:   
$\rho_\chi = \rho_s (r/r_s)^{-1} (1 + r/r_s)^{-2}$, 
where both the scale radius $r_s$ and the scale density $\rho_s$ depend 
on the concentration parameter $c$ which is a function of mass, redshift 
and the cosmological model.   
Since we aim at exploring the most optimistic scenario for neutralino annihilation
we do not consider here the results of some recent high resolution N-body experiments
\cite{nfw:04,merr:05} which suggest that halo density profiles near the center 
have a continuously varying logarithmic slope, and thus are significantly
shallower than the cuspy NFW profile.
In this work we assume the flat, $\Lambda$CDM ``concordance'' model  
and use the numerical routines provided by \cite{ens:01} to compute $c$.   
The smallest halos of  mass $10^{-6} \msun$ turn out to have 
$c \sim {\cal O} (40)$ at $z = 0$. 

The previous assumption allows one to specify the number density of 
subhalos per unit mass at a distance $r$ from the Galactic Center (GC):   
\begin{equation}  
\rho_{sh}(M,r) = A M^{-2} \frac{\theta (r - r_{min}(M))}{(r/r_s^{MW}) 
(1 + r/r_s^{MW})^{2}} \msun^{-1} \kpc^{-3}  
\label{rho}  
\end{equation}  
We set A such that 10\% of the MW mass ($M_{\rm MW}=10^{12} \msun$)  
is distributed in subhalos with masses greater than $10^{7} \msun$  
to match the results of \cite{reed:03}.  
As a result about 53\% of the dark mass within our galaxy is not smoothly   
distributed but contained within $\sim 1.5 \times 10^{16}$ subhalos   
with masses larger than $10^{-6} \msun$, corresponding to   
$\sim 100 \pc^{-3}$ halos in the solar neighborhood.   
  
The $\theta$ term in Eq. \ref{rho} takes into account the   
effect of gravitational tides which, according to the Roche criterion,  
disrupt all halos within $r_{min}(M)$ from the GC \cite{Hayashi:02}
in an orbital period. At  $z=0$ $r_{min}$ is an increasing function of  
the subhalo mass implying that no subhalos survive within  
$r_{min}(10^{-6} \msun) \sim 200 \pc$.  
  
In this work we wish to predict the annihilation signal   
expected in  a $100^o \times 100^o$ field of view (f.o.v.) with an 
angular resolution of $\Delta \Omega=10^{-5} \sr$,    
matching those planned for the GLAST experiment, in both the GC and  
the anticenter (AC) directions.  
This requires to evaluate the line--of--sight integral:  
\begin{equation}  
\Phi^{\rm cosmo}(\psi,\theta) = \int_{\Delta \Omega (\psi,\theta)}  
d \Omega' \int_{\rm l.o.s} \rho_{\chi}^2 (r(\lambda,\psi')) 
d\lambda(r,\psi'),  
\label{flussocosmoMW}  
\end{equation}  
where $\psi$ is the angle--of--view from the GC, $\theta$ is the angular 
resolution of the detector, $\rho_\chi(r)$ is the mass density.  
The distance from the GC $r$ is $r = \sqrt{\lambda^2 + 
\rsun^2 - 2 \lambda \rsun \cos \psi}$, $\lambda$ is the distance from 
the observer and $\rsun=8.5 \kpc$ is that of the Sun from the GC.  
  
We evaluate of Eq. \ref{flussocosmoMW} in two   
steps. First, we numerically integrate Eq. \ref{flussocosmoMW} in which 
$\rho_\chi(r)$ is the sum of NFW profiles corresponding to all subhalos  
distributed according to Eq. \ref{rho} along the l.o.s..  
This gives the average subhalo contribution to the Galactic 
annihilation flux within $10^{-5} \sr$ along the direction $(\psi,\theta)$.  
We found that the average contribution to $\Phi^{\rm cosmo}$,  
in units of $\GeV^2 \cm^{-6} \kpc \sr$, of subhalos in the range 
$10^{-6} \msun < M_{sh} < 10^{6} \msun$ is $4 \times 10^{-5}$ at the GC,  
then slowly decreases to $10^{-5}$ at $\psi = 50^o$ and 
to $4 \times 10^{-6}$ at larger angles.  
For $\phi > 25^o$ it dominates over the Galactic foreground 
accounting for neutralino annihilation in the smooth Galactic halo 
of $4.7 \times 10^{11} \msun$.  
  
To estimate the variance to the average flux we use Eq. \ref{flussocosmoMW}  
to generate several independent Monte Carlo realizations  
of the closest and brightest minihalos.  
For each subhalo mass, we only consider objects which are close enough to   
guarantee $\Phi^{\rm cosmo} > 10^{-6} \GeV^2 \cm^{-6} \kpc \sr$.  
If no such halo exists then we still Monte Carlo generate the 100 closest   
objects in that mass range, since we expect the bigger fluxes to come from the
closer halos.     

\begin{figure}[t]  
\vspace{-25pt}  
\includegraphics[height=8cm,width=8cm]{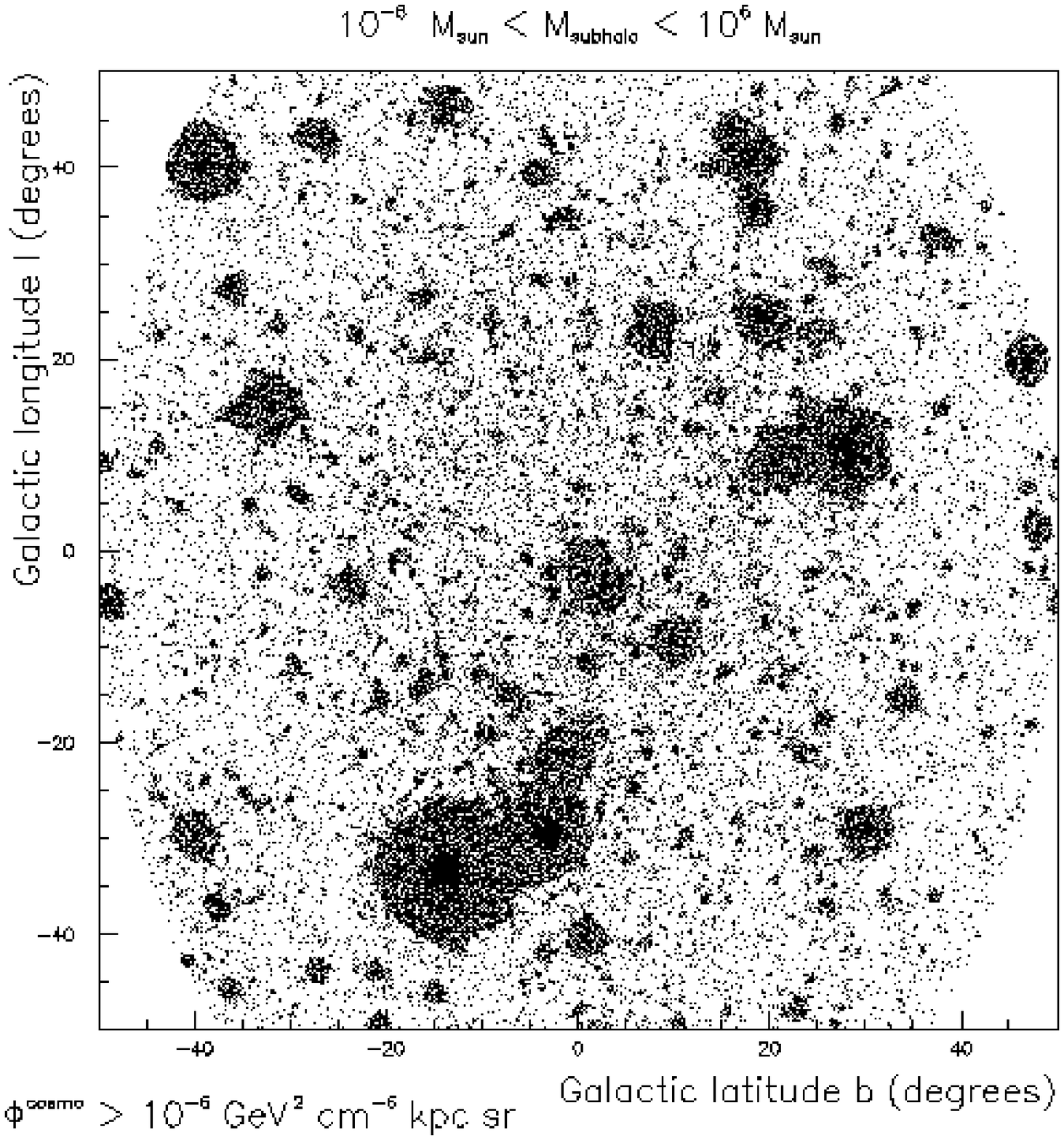}  
\vspace{-5pt}
\caption{Sky map of the annihilation signal from nearby subhalos
of masses $[10^{-6}, 10^{6} \msun]$
contributing more than $\Phi^{\rm cosmo} = 10^{-6} \GeV^2 \cm^{-6} \kpc \sr$
in a $100^o \times 100^o$ f.o.v. centered on the GC in one of our   
Monte Carlo realization.    
}
\label{fig1}  
\end{figure}

The contribution of nearby structures to the annihilation flux   
can be appreciated from Fig.\ref{fig1} which shows the  
sky distribution of the annihilation signal in one of our Monte Carlo   
realization in the direction of the GC.   
    
To compute the total annihilation signal from all sub-Galactic halos 
we have followed the same procedure as \cite{pb:04} and have  
generated several Monte Carlo realizations of subhalos 
with  $10^{6} \msun < M_{sh} < 10^{10} \msun$ to compute their 
contribution to the total flux.
  
\begin{figure}[t]  
\vspace{-25pt}  
\includegraphics[height=7.5cm,width=8cm]{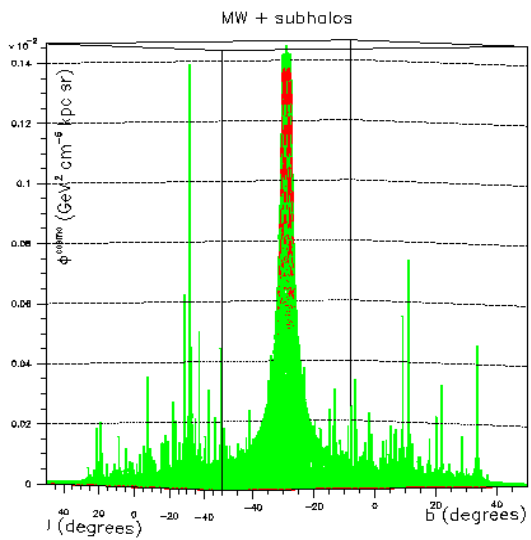}  
\vspace{-5pt}
\caption{3D view of the total $\Phi^{\rm cosmo}$ contributed by subhalos  
and the Galactic foreground, in a $100^o \times 100^o$ f.o.v. centered   
on the GC. The Galactic foreground is cut  
at the subhalos' level.}  
\label{fig2}  
\end{figure}

\begin{figure}[t]  
\vspace{-25pt}  
\includegraphics[height=8cm,width=8cm]{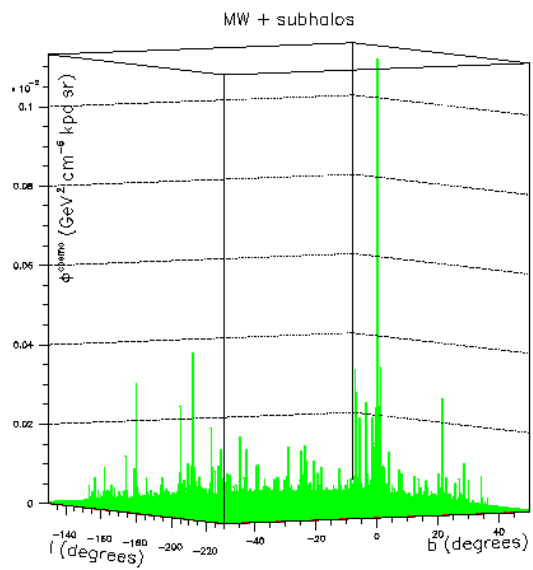}  
\vspace{-5pt}
\caption{ Same as Fig.\ref{fig2} in the direction of the AC. 
}  
\label{fig3}  
\end{figure}

Figs. \ref{fig2} and \ref{fig3} show the expected contribution to  
$\Phi^{\rm cosmo}$ accounting for both subhalos and Galactic foreground  
in the direction of the GC and the AC, respectively.  
As expected, the Galactic foreground is negligible
in Fig. \ref{fig3}, while it is dominating around the GC in  Fig. \ref{fig2}.  
Note that the very prominent peak due to the Galactic foreground at 
the GC position has been artificially truncated to appreciate the 
subhalos contribution.  
We estimate that $\Phi^{\rm cosmo} = 0.04 \GeV^2 \cm^{-6} \kpc \sr$ 
at the GC.   

Figs. \ref{fig2} and \ref{fig3} also show that even the biggest contribution  
to $\Phi^{\rm cosmo}$ from a subhalo is well below the ${\cal O} (1)$
level which, given our best value of $\Phi^{\rm SUSY}$, would be    
required for detection in the GLAST experiment.  In fact, in no
Monte Carlo realizations we have found that the subhalos contribution 
to  $\Phi^{\rm cosmo}$ exceed the value of ${\cal O} (10^{-3})$.
  
To make this statement more quantitative and to also account for the 
possibility of detecting the $\gamma$-ray annihilation line we have   
evaluated the sensitivity of GLAST to the differential spectrum  
of photons expected from DM annihilation, convolved with its energy  
resolution $\Delta E$=10\%. We define the sensitivity $\sigma (\Delta E)$ as  
 $n_{DM}(\Delta E)/\sqrt{n_{\rm bkg}(\Delta E)}$ =   
\begin{equation}  
= \frac{\sqrt{T_\delta} \epsilon_{\Delta \Omega}}{\sqrt{ \Delta \Omega}}  
\frac{\int_{\Delta E} A^{\rm eff}_\gamma (E,\theta) [d\phi^{\rm DM}_\gamma/dE  
d\Omega] dE d\Omega}{\sqrt{\int_{\Delta E} \sum_{\rm bkg} A^{\rm eff}_{\rm  
bkg}(E,\theta) [d\phi_{bkg}/dEd\Omega] dE d\Omega}}.   
\label{sensitivity} \nonumber  
\end{equation}  
where $T_\delta$ defines the effective observation time,  
$\epsilon_{\Delta \Omega} = 0.7$ is the fraction of signal  
events within the optimal solid angle $\Delta \Omega$ corresponding to  
the angular resolution of $0.1^{\circ}$, and $A^{\rm eff}= 10^4 \cm^2$  
is the effective detection area.  
We optimistically assume that the photon and charged particles'   
detection efficiencies $\epsilon_{\gamma}$ and $\epsilon_{\rm ch}$ are 100\%,  
thus we only consider galactic and extragalactic $\gamma$-ray background  
as it is extrapolated by EGRET data at lower energies \cite{Bergstrom:98}.   
We simulated a continuous 5 years observation of a pixel   
of our grid located at $\psi = 55^o$, assuming that in that pixel 
the value of $\Phi^{\rm cosmo}$ is the largest value found among all our 
Monte Carlo realizations.

\begin{figure}[t]  
\vspace{-25pt}  
\includegraphics[height=8cm,width=8cm]{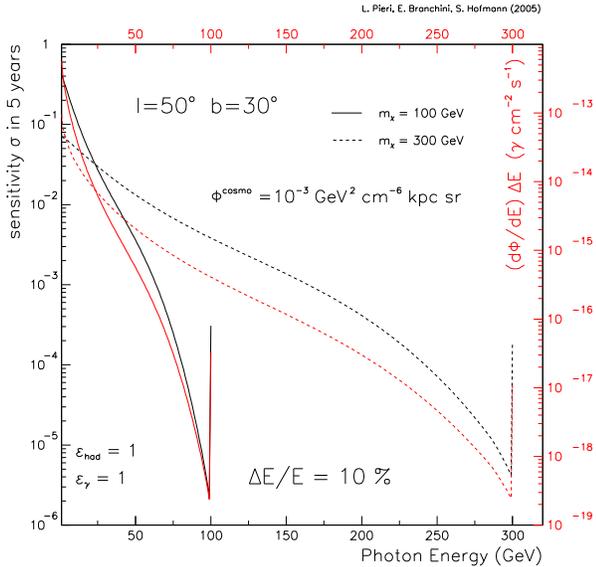}  
\vspace{-5pt}  
\caption{  
Left y axis (black): Experimental sensitivity of GLAST to the photon flux  
from DM annihilation coming from a subhalo at $\psi \sim 55^o$, with   
$\Phi^{\rm cosmo} = 10^{-3} \GeV^2 \cm^{-6} \kpc \sr$ for $m_\chi = 100 \GeV$  
(solid line) and 300 GeV (dashed-line).  
Right y axis (red): Expected flux from the same subhalo as observed  
by GLAST with an energy resolution of 10\%.}  
\label{fig4}  
\end{figure}

We chose a neutralino mass of 100 and 300 GeV, whose  $\gamma$-lines would in
principle be observable with GLAST.
The results are shown in Fig.\ref{fig4}. The black curves, corresponding  
to the left y-axis, show the sensitivity $\sigma (\Delta E)$ of GLAST  
as a function of E, for two different values of the neutralino mass. 
The red curves, to be read on the right y-axis, show the expected flux 
from the same pixel. It is evident that GLAST would hardly detect 
neither continuum flux nor $\gamma$-lines from Galactic subhalos.

Since it seems implausible to significantly increase $\Phi^{\rm SUSY}$,  
our results confirms those of \cite{cm:00,pb:04,pmp:04,otn:05}
show that currently planned, satellite-borne experiments   
such as GLAST will not be able to detect the annihilation signal  
produced by Galactic subhalos.
This is due to the fact that, as pointed out by \cite{sto:03},
the total annihilation flux is dominated by  massive Galactic 
subhalos rather than by minihalos.
A result which makes our conclusion insensitive to variations in the
low-mass density cut-off, in agreement with \cite{savvas:04}.
  
All plausible cosmological and astrophysical effects  
like the existence of a central core in the halo density profile,  
the various dynamical disturbances that reduce the subhalo survival   
probability, the possible existence of a cutoff scale $\sim 0.1 k_{\rm cut}$  
found \cite{lz:05} in the CDM power spectrum, generated by 
acoustic oscillations with wavelength comparable to the size of the horizon
at kinetic decoupling, would decrease the expected annihilation signal, 
making it our conclusions for subhalos more pessimistic.
Unless, of course, one adopts steeper density profiles 
that, however, are not supported by recent numerical experiments.
Note that the $\gamma$-ray luminosity of our minihalos, which is designed 
to match the prediction of \cite{dms:05}, is significantly smaller 
than that of \cite{otn:05}; 
a difference that traces back to the large internal 
density of their minihalos.

The presence of a minihalo population like that considered in 
this work also enhance the $\gamma$-ray flux from nearby 
objects like M31. We estimate that, for a NFW profile, the total 
$\Phi^{\rm cosmo}$ due to both the smooth and the subhalo contribution 
will not exceed the value of $5 \times 10^{-4}$, making it 
impossible to detect.

Recently, the HESS telescope has announced the serendipitous discovery of an 
unidentified extended TeV $\gamma$-ray source, with a total flux above 
380 GeV of $ \sim 10^{-11} \cm^{-2} \sec^{-1}$ \cite{hess:05}.
If explained in terms of the annihilation of ${\cal O} (10 \TeV)$ 
neutralinos, such a large flux could only be 
accounted for by advocating a central density  $r^{-1.5}$, steeper than the NFW model.
A prominent central spike might form around a supermassive black hole
\cite{bertone:05,bertone2:05} that, according to hierarchical models of galaxy formation
\cite{volonteri:03}, could reside at the center of both massive Galactic and sub-Galactic halos.

We acknowledge fruitful discussions with L. Bergstr\"om, D. Els\"asser, P. Faccioli, A.M. Green, A. Lionetto, 
K. Mannheim, B. Moore, A. Morselli and  D.J. Schwarz.

\bibliography{prl0}

\end{document}